\documentclass[pra,aps,twocolumn,showpacs,amssymb]{revtex4} 
\usepackage{epsfig} 
\usepackage{amscd} 
\usepackage{bm} 
\newcommand{\ket}[1]{|#1\rangle} 
\newcommand{\bra}[1]{\langle #1|} 
\newcommand{\Tr}{\text{Tr}} 
 
\newcommand{\mod}{\text{\,mod\,}}%{\mathtt{\%}} 
\begin{document} 
\title{Off-diagonal generalization of the mixed state geometric phase} 
\author{Stefan Filipp$^{1}$\footnote{Electronic address: 
sfilipp@ati.ac.at} and Erik Sj\"{o}qvist$^{2}$\footnote{ 
Electronic address: erik.sjoqvist@kvac.uu.se}} 
\affiliation{$^{1}$ Atominstitut der \"{O}sterreichischen 
Universit\"{a}ten, Stadionallee 2, A-1020 Vienna, Austria \\ 
$^{2}$Department of Quantum Chemistry, Uppsala University, 
Box 518, Se-751 20 Uppsala, Sweden} 
\date{\today} 
\begin{abstract} 
The concept of off-diagonal geometric phases for mixed quantal states
in unitary evolution is developed. We show that these phases arise
from three basic ideas: (1) fulfillment of quantum parallel transport
of a complete basis, (2) a concept of mixed state orthogonality
adapted to unitary evolution, and (3) a normalization condition. We
provide a method for computing the off-diagonal mixed state phases to
any order for unitarities that divide the parallel transported basis
of Hilbert space into two parts: one part where each basis vector
undergoes cyclic evolution and one part where all basis vectors are
permuted among each other. We also demonstrate a purification based
experimental procedure for the two lowest order mixed state phases and
consider a physical scenario for a full characterization of the qubit
mixed state geometric phases in terms of polarization-entangled photon
pairs. An alternative second order off-diagonal mixed state geometric
phase, which can be tested in single-particle experiments, is
proposed.
\end{abstract} 
\maketitle 
\section{Introduction} 
The concept of geometric phase, anticipated by Pancharatnam 
\cite{pancharatnam56} in his study of interference of classical 
light in distinct states of polarization and developed by Berry
\cite{berry84} for cyclic adiabatic quantal evolution, has been
generalized in several steps. Aharonov and Anandan \cite{aharonov87}
considered the cyclic nonadiabatic case and pointed out that the
geometric phase is due to the curvature of the space of pure quantal
states. Samuel and Bhandari \cite{samuel88} provided a general setting
for the geometric phase so as to cover noncyclic evolution and
sequential projection measurements. Another line of development has
been the extension of the geometric phase to the mixed state
case. Uhlmann \cite{uhlmann86} was probably first to address this
issue in the mathematical context of purification and Sj\"{o}qvist
{\it et al.} \cite{sjoqvist00} discovered a mixed state geometric
phase for nondegenerate density operators in noncyclic unitary
evolution in interferometry. The parallel transport conditions in
Refs. \cite{uhlmann86,sjoqvist00} have been shown to lead to
generically different interference effects \cite{slater02} and the
mixed state concept in Ref. \cite{sjoqvist00} has been extended to the
case of completely positive maps \cite{ericsson03} as well as to
degenerate density operators \cite{singh03}. An experimental test of
the proposed mixed state geometric phase in Ref. \cite{sjoqvist00} in
the qubit case has been reported recently \cite{du03}, using nuclear
magnetic resonance technique.

The noncyclic geometric phases in Refs. \cite{samuel88,sjoqvist00} 
become undefined when the interfering states are orthogonal and 
the interference visibility vanishes. This leads to an interesting 
nodal point structure in the experimental parameter space that 
could be monitored in a history-dependent manner 
\cite{bhandari91,bhandari02}. The existence of nodal points also 
led Manini and Pistolesi \cite{manini00} to introduce the off-diagonal
geometric phases for pure states in adiabatic evolution. These phases
may carry interference information at the nodal points of the standard
geometric phase. The adiabaticity assumption of Ref. \cite{manini00}
was subsequently removed by Mukunda {\it et al.} \cite{mukunda02} and
the second order off-diagonal pure state geometric phase was verified
by Hasegawa {\it et al.} \cite{hasegawa01,hasegawa02} for neutron
spin. More recently, the off-diagonal generalization of the mixed
state phase in Ref. \cite{sjoqvist00} for parallel transporting 
unitarities has been proposed \cite{filipp03} and extended to 
general unitarities \cite{sjoqvist03} by the present authors.

In this paper, we wish to elaborate further on the concept in
Ref. \cite{filipp03}. Our first concern is to develop systematically
the off-diagonal mixed state geometric phase in unitary evolution,
filling in some important conceptual gaps of Ref. \cite{filipp03}. The
theory of parallel transport of a complete basis is considered. Under
such a parallel transport, it is argued that the mixed state geometric
phase follows naturally from a concept of orthogonality between
density operators adapted to unitary evolution and a certain
normalization condition. Our second concern is to provide a method for
computing mixed state geometric phase factors to any order for the
important class of unitarities under which the parallel transported
basis of Hilbert space is divided into two parts: one part where each
basis vector undergoes cyclic evolution and one part where all basis
vectors are permuted among each other. The experimental realization of
the two lowest order mixed state phases is discussed, using the idea
of purification of the involved mixed states by attaching an ancilla
system entangled to the considered system. We further develop the
two-photon experiment proposed in Ref. \cite{filipp03} so as to
include also the lowest order mixed state phase. Finally, as an
attempt to avoid the apparent need for an ancilla system, we consider
an alternative off-diagonal mixed state geometric phase conceptually 
based upon the neutron experiment in Refs. \cite{hasegawa01,hasegawa02}.

\section{Basic ideas} 
\subsection{Quantum parallel transport} 
Consider a continuous one-parameter family $\{ U(s),s\in [s_0,s_1] 
\big| U(s_0) = I \}$ of unitarities that map any initial complete 
orthonormal basis $\{ \ket{\psi_k } \}$ of a Hilbert space ${\cal H}$
of dimension $N$ to a continuous set of complete orthonormal
bases $\{ \ket{\psi_k (s)} , s\in [s_0,s_1] \}$ of the same ${\cal
H}$. In the context of off-diagonal phases, it proves useful to
consider parallel transport of such a complete set of orthonormal 
pure state vectors.

Parallel transport in terms of the initial basis $\{ \ket{\psi_k } \}$ 
may be formulated as follows. Let $J(s)$ be the Hamiltonian operator 
in the Heisenberg representation. For the corresponding evolution 
operator $U(s)$ we have ($\hbar = 1$ from now on) 
\begin{equation} 
i\dot{U} (s) = U(s) J(s) ,
\end{equation}
the formal solution of which reads 
\begin{equation} 
U(s) = {\cal P}^{-1} \exp \left( -i \int_{s_0}^{s} J(s') ds' \right) . 
\end{equation} 
Here, ${\cal P}^{-1}$ is inverse path ordering, i.e., ordering
increasing $s$ from left to right in each term of series expansion
of the evolution operator. $U(s)$ is said to parallel transport 
the initial basis $\{ \ket{\psi_k} \}$ if the local accumulation 
of phase along the unitary path vanishes for each $\ket{\psi_k}$, 
which amounts to $\bra{\psi_k} U^{\dagger} (s) \dot{U} (s) 
\ket{\psi_k} = 0$, i.e., 
\begin{equation} 
\bra{\psi_k} J(s) \ket{\psi_k} = 0, \ \forall k . 
\label{eq:pcj} 
\end{equation} 
Thus, the one-parameter family of Hermitian generators $J(s)$ has to 
be off-diagonal in the parallel transported initial basis and therefore 
traceless in any basis. In other words, $U \in$ SU(N) is a necessary 
condition for $U$ being parallel transporting a complete basis. 
The converse does not hold: there are SU(N) transformations that 
contain generators that have nonvanishing diagonal elements in the 
$\{ \ket{\psi_k} \}$ basis.

We may equally well formulate the parallel transport conditions 
in terms of the instantaneous basis $\{ \ket{\psi_k (s)} \}$. 
Here, we have $\bra{\psi_k (s)} \dot{U} (s) U^{\dagger} (s) 
\ket{\psi_k (s)} = 0$, which entails 
\begin{equation} 
\bra{\psi_k (s)} H(s) \ket{\psi_k (s)} = 0, \ \forall k , 
\label{eq:pch} 
\end{equation} 
where $H(s) = U(s) J(s) U^{\dagger} (s)$ is the Hamiltonian 
operator in the Schr\"{o}dinger picture. Thus, $H(s)$ has to 
be off-diagonal in the instantaneous parallel transported basis, 
which is consistent with $U \in$ SU(N). 

Any parallel transporting unitarity is denoted by $U^{\parallel}$ in
the following. Moreover, an initial (instantaneous) nondegenerate 
\cite{remark1} density operator whose eigenvectors coincide with 
the basis $\{ \ket{\psi_k} \}$ ($\{ \ket{\psi_k (s)} \}$) is said 
to be the parallel transported by $U^{\parallel}$ fulfilling 
Eqs. (\ref{eq:pcj}) and (\ref{eq:pch}).

For $\dim {\cal H} = N$, any $J(s)$ fulfilling the parallel transport
conditions may be written in terms of $N^2-N$ linearly independent
off-diagonal generators in the basis $\{ \ket{\psi_k} \}$.  As an
example, consider the qubit case $N=2$. Here, we expect $J(s)$ be
dependent upon two of the Pauli operators. Let
$\{ \ket{\psi_1},\ket{\psi_2} \}$ be a parallel transported basis. 
Then $J(s) = a(s) \sigma_x + b(s) \sigma_y$ with $\sigma_x = 
\ket{\psi_1} \bra{\psi_2} + \ket{\psi_2} \bra{\psi_1}$,  
$\sigma_y = -i\ket{\psi_1} \bra{\psi_2} + i\ket{\psi_2} \bra{\psi_1}$,
and $a(s),b(s)$ being scalar functions of $s$. Any density operator
$\rho$ of the form $\rho = \lambda_1 \ket{\psi_1}
\bra{\psi_1} + \lambda_2 \ket{\psi_2} \bra{\psi_2}$, $\lambda_1 \neq 
\lambda_2$ is parallel transported by the corresponding unitarity. 
The same holds true for $H(s)$ by making the replacement 
$\{ \ket{\psi_1},\ket{\psi_2} \} \rightarrow 
\{ \ket{\psi_1 (s)},\ket{\psi_2 (s)} \}$.

\subsection{Orthogonality} 
Two pure state vectors are orthogonal if their scalar product vanishes. On
the other hand, any useful scalar product between density operators does 
not have this simple property and the concept of orthogonality becomes
less straightforward in the mixed state case. Instead, for a given
density operator $\rho$, one may take another density operator $\rho'$
to be orthogonal to $\rho$ if it yields minimum of the Hilbert-Schmidt
product $\Tr [\rho \rho']$ or the Bures fidelity
\cite{bures69} ${\cal F}_B [\rho ,\rho' ] = \big[\Tr
\sqrt{\sqrt{\rho}\rho'
\sqrt{\rho}} \big]^2$, the latter being a worst case measure 
of distinguishability between $\rho$ and $\rho'$ \cite{jozsa94}.
Here, we take a third approach to the concept of orthogonality adapted
to unitarily connected density operators. The idea is to say that
$\rho$ and $\rho' = U\rho U^{\dagger}$ are orthogonal whenever they  
cannot interfere in the sense of Ref. \cite{sjoqvist00}. 

To develop this idea in detail, let us first suppose $\ket{\psi}$ and
$\ket{\varphi}$ are Hilbert space representatives of two arbitrary
pure quantal states $\psi$ and $\varphi$, and assume
further that $\ket{\psi}$ is exposed to the variable U(1) shift
$e^{i\chi}$.  The resulting interference pattern obtained in
superposition is determined by the intensity profile 
\cite{wagh95}
\begin{equation} 
{\cal I} \propto \Big| e^{i\chi} \ket{\psi} + \ket{\varphi} 
\Big|^{2} = 2 + 2 \big| \langle \psi | \varphi \rangle \big| 
\cos \big[ \chi - \arg \langle \psi | \varphi \rangle \big]  
\label{eq:pureinterfer} 
\end{equation} 
that oscillates as a function of $\chi$. The key point here is 
to note that $\psi$ and $\varphi$ are orthogonal if and only if 
${\cal I}$ is independent of $\chi$ so that the interference 
oscillations disappear. 
 
This feature translates naturally to the mixed state case. 
Consider a pair of isospectral nondegenerate density operators 
\begin{equation} 
\rho_{\psi} = 
\sum_{k} \lambda_{k} |\psi_{k} \rangle \langle \psi_{k}| , \ \ 
\rho_{\varphi} = 
\sum_{k} \lambda_{k} |\varphi_{k} \rangle \langle \varphi_{k}| , 
\end{equation} 
where each $|\varphi_{k}\rangle = U|\psi_{k} \rangle$ for some
unitarity $U$. Each such orthonormal pure state component of the
density operator contributes to the interference according to
Eq. (\ref{eq:pureinterfer}). Thus, the total intensity profile 
becomes \cite{sjoqvist00}
\begin{eqnarray} 
{\cal I} & \propto & 
\sum_{k} \lambda_{k} \Big| e^{i\chi}|\psi_{k}\rangle + 
|\varphi_{k}\rangle \Big|^{2} 
\nonumber \\ 
 & = & 2 + 2 \sum_{k} \lambda_{k} 
\big| \langle \psi_{k} | \varphi_{k} \rangle \big| 
\cos \big[ \chi - \arg \langle 
\psi_{k} | \varphi_{k} \rangle \big] , 
\end{eqnarray} 
where we have used that the $\lambda$'s sum up to unity. 
Following the above pure state case, we say that $\rho_{\psi}$ and 
$\rho_{\varphi}$ are orthogonal if and only if ${\cal I}$ is independent 
of $\chi$ for all Hilbert space representatives $\{ \ket{\psi_k} \}$ 
and $\{ \ket{\varphi_k} \}$ of the eigenstates of $\rho_{\psi}$ and 
$\rho_{\varphi}$, respectively. It follows that $\rho_{\psi}$ and 
$\rho_{\varphi}$ are orthogonal if and only if 
$\langle \psi_k\ket{\varphi_k} = 0, \ \forall k$. 
 
For an $N$ dimensional Hilbert space ${\cal H}$, we may 
generate a set of $N$ mutually orthogonal density operators 
as follows. Assume $\rho_{1} \ket{\psi_k} = \lambda_k \ket{\psi_k}$  
is nondegenerate and introduce a unitary operator $U_g$ such that 
$\ket{\psi_n} = \big( U_g \big)^{n-1} \ket{\psi_1}, \ n=1,\ldots ,N$. 
Thus, we may write 
\begin{equation} 
\label{eq:Uperm} 
U_g = \ket{\psi_1} \bra{\psi_N} + \ket{\psi_N} \bra{\psi_{N-1}} + 
\ldots \ket{\psi_2} \bra{\psi_1}  
\end{equation} 
and it follows that 
\begin{equation} 
\rho_{n} = \big( U_g \big)^{n-1} \rho_{1} \big( U_g^{\dagger} 
\big)^{n-1} , \ n=1,\ldots , N 
\end{equation} 
is a set of mutually orthogonal density operators. Explicitly, 
this entails that 
\begin{eqnarray} 
\label{eq:rhoort} 
\rho_{1} & = & \lambda_{1} \ket{\psi_1} \bra{\psi_1} + 
\lambda_{2} \ket{\psi_2} \bra{\psi_2} + \ldots + 
\lambda_{N} \ket{\psi_N} \bra{\psi_N} , 
\nonumber \\ 
\rho_{2} & = & \lambda_{1} \ket{\psi_2} \bra{\psi_2} + 
\lambda_{2} \ket{\psi_3} \bra{\psi_3} + \ldots + 
\lambda_{N} \ket{\psi_1} \bra{\psi_1} , 
\nonumber \\ 
 & \ldots , & 
\nonumber \\ 
\rho_{N} & = & \lambda_{1} \ket{\psi_N} \bra{\psi_N} + 
\lambda_{2} \ket{\psi_1} \bra{\psi_1} + \ldots 
\nonumber \\ 
 & &  + \lambda_{N} \ket{\psi_{N-1}} \bra{\psi_{N-1}} . 
\label{eq:Nstates} 
\end{eqnarray} 
Notice here that different sets of mutually orthogonal mixed states 
may be generated by permuting the $\psi_n$'s in $U_g$.  

\subsection{Consistency and normalization} 
The final step towards the concept of off-diagonal mixed state 
geometric phase, is to determine how the mutually orthogonal 
density operators should appear in the trace. This may be resolved 
as follows. 

We first notice that the Manini-Pistolesi expression  
\cite{manini00} may be written in terms of pure state projectors 
$P_{j_k} = \ket{\psi_{j_k}} \bra{\psi_{j_k}}$ as 
\begin{equation} 
\gamma_{P_{j_1} P_{j_2} \ldots P_{j_l}}^{(l)} 
\equiv \Phi \big[ \Tr \big( U^{\parallel} P_{j_1} 
U^{\parallel} P_{j_2} \ldots U^{\parallel} P_{j_l} \big) \big] ,  
\label{eq:man} 
\end{equation} 
where $\Phi[z]=z/|z|$ for any complex number $z$. We propose to
replace each of these projectors with $F^{(l)} (\rho_{j_k})$, where,
for reason of permutation symmetry of the indexes $j_1,j_2, \ldots
,j_l$, the form of the function $F^{(l)}$ may only depend on $l$. To
assure consistency with Ref. \cite{manini00} we further require 
that $F^{(l)} (\rho_{j_k}) \rightarrow P_{j_k}$ in the pure state 
limit. We take the simplest nontrivial choice fulfilling this 
requirement, which is $F^{(l)} (\rho_{j_k}) = \rho_{j_k}^{p/q}$ 
\cite{remark2}, $p=p(l)$ and $q=q(l)$ integers \cite{remark3}. 

Next, from $\big( P_k \big)^l = P_k$, we obtain the normalization 
condition
\begin{eqnarray} 
 & & \Tr \big( U_g^{\dagger} P_k U_g^{\dagger} P_{(k+1) \mod N} 
\ldots U_g^{\dagger} P_{(k+l) \mod N} \big) 
\nonumber \\ 
 & & = \Tr \big( \big( U_g^{\dagger} \big)^l P_k \big) = 
\delta_{lN} , \ \forall k \in [1,N] , 
\end{eqnarray} 
where we have used $U_g^{\dagger}$ defined in Eq.~(\ref{eq:Uperm}), 
$P_{(k+n) \mod N} = \big( U_g \big)^{n-1} P_{k} \big( 
U_g^{\dagger} \big)^{n-1}$ and $\big( U_g^{\dagger} \big)^N = I$. 
We propose to demand that this normalization structure is preserved 
in the mixed state case. After the replacement $P_{j_k} \rightarrow 
\rho_{j_k}^{p/q}$, we similarly have
\begin{eqnarray} 
 & & \Tr \big( U_g^{\dagger} \rho_k^{p/q} U_g^{\dagger} 
\rho_{(k+1) \mod N}^{p/q} \ldots U_g^{\dagger} 
\rho_{(k+l) \mod N}^{p/q} \big) 
\nonumber \\ 
 & & = \Tr \big( \big( U_g^{\dagger} \big)^l \rho_k^{lp/q} \big) = 
\delta_{lN} \Tr \big( \rho_k^{lp/q} \big) , \ 
\forall k \in [1,N] , 
\nonumber \\ 
\end{eqnarray}  
where we have used that $\big( U_g \rho U_g^{\dagger} \big)^{p/q} =
U_g \rho^{p/q} U_g^{\dagger}$. Thus, only $p(N)=1$ and $q(N)=N$
assures the desired kind of normalization in the mixed state case. 
Since $p$ and $q$ are functions of $l$ only, it follows that
$p=1$ and $q=l$.

This choice may also be understood from the following simple 
convergence arguments in the $N\rightarrow \infty$ case. 
$\rho^{lp/q}$ typically involves factors of the form $\lambda^{lp/q}$, 
$0 \leq \lambda \leq 1$. If $lp/q < 1$ ($lp/q > 1$) then the trace 
diverges (goes to zero) when $l \rightarrow \infty$. Thus, only for 
$lp/q = 1$ the $N \rightarrow \infty$ limit is finite and well defined. 

\section{Off-diagonal mixed state geometric phase} 
We are now ready to state our main result. The off-diagonal mixed
state phase for an ordered set of $l\leq N$ mutually orthogonal
nondegenerate density operators $\rho_{j_k}$, $k=1, \ldots ,l$,
parallel transported by $U^{\parallel}$ is naturally given by
\begin{equation} 
\gamma_{\rho_{j_1}\rho_{j_2}\ldots\rho_{j_l}}^{(l)} 
\equiv \Phi \big[ \Tr \big( U^{\parallel} \sqrt[l]{\rho_{j_1}} 
U^{\parallel} \sqrt[l]{\rho_{j_2}} \ldots 
U^{\parallel} \sqrt[l]{\rho_{j_l}} \big) \big] . 
\label{eq:genoffdiag} 
\end{equation} 
This is manifestly gauge invariant and independent of cyclic
permutations of the indexes $j_{1},j_{2}, \ldots ,j_{l}$. By
construction it reduces to Eq. (\ref{eq:man}) in the limit of pure
states.
 
The mixed state geometric phase factor  
\begin{equation} 
\gamma_{\rho_{j_1}}^{(1)} = 
\Phi \big[ \Tr \big( U^{\parallel} \rho_{j_1} \big) \big] 
\label{eq:mixedl1} 
\end{equation} 
proposed in Ref. \cite{sjoqvist00} may be seen as a natural 
consequence of this general framework if we put $l=1$. In Sec. V 
we propose experimental realization of this first ($l=1$) and 
second order ($l=2$) phases, the latter being defined by   
\begin{equation} 
\gamma_{\rho_{j_1} \rho_{j_2}}^{(2)} = 
\Phi \big[ \Tr \big( U^{\parallel} \sqrt{\rho_{j_1}} 
U^{\parallel} \sqrt{\rho_{j_2}} \big) \big] , 
\label{eq:mixedl2} 
\end{equation} 
in polarization-entangled two-photon interferometry. 
 
\section{Computation of off-diagonal mixed state phases} 
In the qubit case $N=2$, consider the unitarity 
\begin{eqnarray} 
U^{\parallel} & = & U_{11}^{\parallel} \ket{\psi_1} \bra{\psi_1} + 
U_{12}^{\parallel} \ket{\psi_1} \bra{\psi_2} + 
U_{21}^{\parallel} \ket{\psi_2} \bra{\psi_1} 
\nonumber \\ 
 & & + U_{22}^{\parallel} \ket{\psi_2} \bra{\psi_2} 
\label{eq:qubitU}
\end{eqnarray} 
that parallel transport some orthonormal basis $\{ \ket{\psi_1}, 
\ket{\psi_2} \}$. The matrix elements of $U^{\parallel}$ fulfill 
$U_{11}^{\parallel}$ $= (U_{22}^{\parallel})^{\ast} = \eta 
e^{-i\Omega /2}$ and $U_{12}^{\parallel} U_{21}^{\parallel} = 
- \det U^{\parallel} + U_{11}^{\parallel} U_{22}^{\parallel} = 
-1+\eta^2$ as $U^{\parallel} \in$ SU(2). Here, $\eta = 
\big| \bra{\psi_1} U^{\parallel} \ket{\psi_1}\big|$ is the 
pure state visibility and $\Omega$ is the solid angle enclosed 
by the path traced out by the basis vectors $\{ \ket{\psi_1}, 
\ket{\psi_2} \}$ and the shortest geodesic connecting its end 
points on the Bloch sphere. 

Now, $U^{\parallel}$ in Eq. (\ref{eq:qubitU}) parallel 
transports the mutually orthogonal density operators 
$\rho_1 = \lambda_1 \ket{\psi_1} \bra{\psi_1} + 
\lambda_2 \ket{\psi_2} \bra{\psi_2}$ and $\rho_2 = 
\lambda_1 \ket{\psi_2} \bra{\psi_2} + \lambda_2 
\ket{\psi_1} \bra{\psi_1}$, for which we obtain 
\begin{eqnarray}  
\Tr \big( U^{\parallel} \rho_{1} \big) & = & 
\Tr \big( U^{\parallel} \rho_{2} \big)^{\ast} 
\nonumber \\ 
 & = &   
\eta \big( \lambda_1 e^{-i\Omega/2} + \lambda_2 e^{i\Omega/2} \big) , 
\nonumber \\ 
\Tr \big( U^{\parallel} \sqrt{\rho_1} U^{\parallel} 
\sqrt{\rho_2} \big) & = & -1 + \eta^2 + 2 \eta^2 
\sqrt{\lambda_1\lambda_2} \cos \Omega 
\nonumber \\ 
 & = & -1 + \eta^2 + 
\eta^{2} \sqrt{{\cal F}_{B} [\rho_1 ,\rho_2 ]} \cos \Omega , 
\nonumber \\  
\label{eq:offqubit} 
\end{eqnarray} 
where we have used the Bures fidelity ${\cal F}_B [\rho_1 ,\rho_2 ] = 
\big[\Tr \sqrt{\sqrt{\rho_1} \rho_2\sqrt{\rho_1}} \, \big]^2 = 
4\lambda_1\lambda_2$. Notice that ${\cal F}_B [\rho_1 ,\rho_2 ] = 0$ 
for pure states and ${\cal F}_B [\rho_1 ,\rho_2 ] = 1$ in the 
maximally mixed state case. 
 
In the nondegenerate mixed state case $\lambda_1 \neq \lambda_2$, 
the $l=1$ phases are indeterminate only for $\eta =0$, for which 
the $l=2$ phase is well-defined since $\Tr \big( U^{\parallel} 
\sqrt{\rho_1} U^{\parallel} \sqrt{\rho_2} \big) = -1$. In the 
degenerate case $\lambda_1 = \lambda_2$, the density operators
$\rho_1$ and $\rho_2$ become identical and spherically symmetric, so
that no specific basis is singled out by the parallel transport
condition and the mixed state geometric phase factors $\gamma^{(1)}$ and
$\gamma^{(2)}$ therefore become undefined.  Still, there is a unique
notion of relative phase in this case with additional nodal points, as
discussed in Ref. \cite{bhandari02}. For a generic $U=e^{-i\delta 
{\bf n} \cdot {\mbox{\boldmath $\sigma$}}}$, 
${\mbox{\boldmath $\sigma$}} = (\sigma_x,\sigma_y,\sigma_z)$ 
being the standard Pauli operators and $|{\bf n}| = 1$, we 
obtain for $l=1$ nodal points at 
$\Tr \big( U \rho_{1} \big) = \Tr \big( U \rho_{1} \big) = 
\cos \delta = 0$ at which $\delta$ we have   $\Tr \big( U 
\sqrt{\rho_1} U \sqrt{\rho_2} \big) = \cos 2\delta = -1$. 
This shows that the $l=1$ and $l=2$ phases never become 
indeterminate simultaneously and thus provide a complete 
phase characterization of the qubit case.
 
\begin{figure}[ht!] 
\begin{center} 
\includegraphics[width=8 cm]{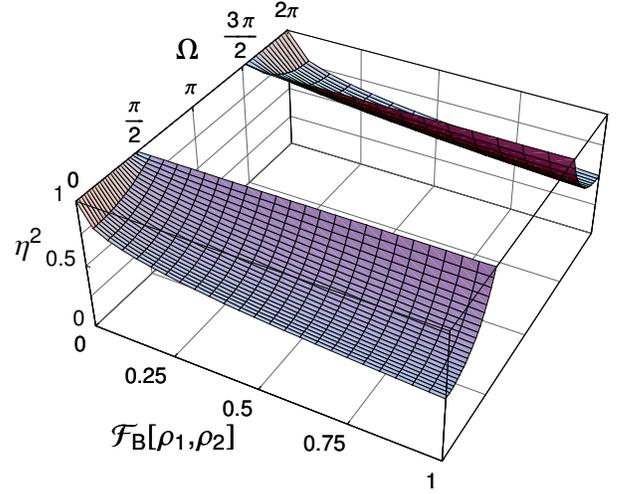} 
\end{center} 
\caption{Nodal surfaces of the off-diagonal mixed state geometric 
phase for a qubit with the solid angle $\Omega$ in
steradians. For Bures fidelity ${\cal F}_{B} =
\big[\Tr \sqrt{\sqrt{\rho_1} \rho_2\sqrt{\rho_1}} \, \big]^2 >0$ 
(mixed states), there are nodes also for paths with pure state 
visibility $\eta = \big| \bra{\psi_1} U^{\parallel} \ket{\psi_1} 
\big|^2 \neq 1$ at various solid angles.}
\label{fig:nodal} 
\end{figure} 
 
The off-diagonal mixed state geometric phase in the qubit 
case has a nontrivial nodal structure that arises due to 
the nonvanishing Bures fidelity. This can be seen by putting 
the left-hand side of Eq. (\ref{eq:offqubit}) to zero and 
solving for $\eta^2$ yielding 
\begin{equation} 
\eta^{2} = \big( 1+\sqrt{{\cal F}_{B} [ \rho_1 ,\rho_2 ]} 
\cos \Omega \big)^{-1} , 
\label{eq:qubitnodal} 
\end{equation} 
which has solutions at $\eta < 1$ for 
${\cal F}_{B} [ \rho_1 ,\rho_2 ] \cos \Omega > 0$. Thus, the 
off-diagonal mixed state geometric phase factor may 
change sign across the nodal surfaces in the parameter space 
$({\cal F}_{B} [ \rho_1 ,\rho_2 ],\eta,\Omega)$ defined by 
the solutions of Eq. (\ref{eq:qubitnodal}), as shown in 
Fig. \ref{fig:nodal}. Thus, the corresponding off-diagonal mixed state 
geometric phase can take both values $0$ and $\pi$, contrary 
to the corresponding pure state phase, which can only be $\pi$. 
 
To generalize the discussion we proceed to arbitrary Hilbert space
dimensions $N$ and provide a method for computing mixed state 
geometric phases to any order $l\leq N$ for unitarities under 
which the parallel transported eigenbasis $\{ \ket{\psi_1}, \ldots, 
\ket{\psi_N} \}$ of the mutually orthogonal $\rho$'s is divided 
into two parts: one part where each basis vector undergoes cyclic 
evolution and one part where all basis vectors are permuted among 
each other. By appropriate labeling of the eigenvectors, such 
unitarities can always be decomposed into the direct sum 
\begin{equation} 
U^{\parallel} = u_p^{\parallel} \oplus u_d^{\parallel} ,  
\end{equation} 
where $u_p^{\parallel}$ permutes $\ket{\psi_1} \rightarrow 
\ket{\psi_m} \rightarrow \ldots \rightarrow \ket{\psi_2} 
\rightarrow \ket{\psi_1}$ and $u_d^{\parallel}$ is diagonal 
in the remaining $N-m$ cyclic eigenvectors. These terms do 
not mix so that one may write 
\begin{eqnarray} 
\Tr \big( U^{\parallel} \sqrt[l]{\rho_{j_1}} \ldots 
U^{\parallel} \sqrt[l]{\rho_{j_l}} \big) = 
{\cal P}_{\rho_{j_1}\ldots\rho_{j_l}}^{(l)} + 
{\cal D}_{\rho_{j_1}\ldots\rho_{j_l}}^{(l)} , 
\end{eqnarray} 
where ${\cal P}_{\rho_{j_1}\ldots\rho_{j_l}}^{(l)} = 
\Tr \big( u_p^{\parallel} \sqrt[l]{\rho_{j_1}} \ldots 
u_p^{\parallel} \sqrt[l]{\rho_{j_l}} \big)$ and 
${\cal D}_{\rho_{j_1}\ldots\rho_{j_l}}^{(l)} = 
\Tr \big( u_d^{\parallel} \sqrt[l]{\rho_{j_1}} 
\ldots u_d^{\parallel} \sqrt[l]{\rho_{j_l}} \big)$. 
 
Turning first to the contribution from the 
diagonal part of $U^{\parallel}$, we have 
\begin{eqnarray} 
{\cal D}_{\rho_{j_1}\ldots\rho_{j_l}}^{(l)} = 
\sum_{k=m+1}^N \big( U_{kk}^{\parallel} \big)^l 
\sqrt[l]{\lambda_{k_1} \ldots \lambda_{k_l}} 
\label{eq:diagonal} 
\end{eqnarray} 
with $U_{kk}^{\parallel}$ the matrix elements of 
$u_{d}^{\parallel}$ in the eigenbasis of the $\rho$'s. 
As the density operators are nondegenerate, it follows that 
all $\lambda_{k_\alpha}$ are different in each term on the 
right-hand side of Eq. (\ref{eq:diagonal}) and ${\cal D}_{\rho_{j_1} 
\ldots\rho_{j_l}}^{(l)}$ must vanish if $l>$ rank of the 
$\rho$'s. Notice here that $\arg U_{kk}^{\parallel}$ is the 
standard cyclic geometric phase for the pure state $\psi_k$. 
 
Considering the contribution from $u_{p}^{\parallel}$ we can establish 
the following. If $l=K\times m$, $K$ integer $\leq N/m$ and $m\geq 2$, 
then 
\begin{eqnarray} 
\label{eq:permutation} 
{\cal P}_{\rho_{j_1}\ldots\rho_{j_l}}^{(l)} =  
\big[ (-1)^{m-1} \det u_p^{\parallel} \big]^K 
f_{\rho_{j_1}\ldots\rho_{j_l}}^{(l)} 
(\lambda_1,\ldots,\lambda_N) ,  
\end{eqnarray} 
where the $f^{(l)}$'s can be written as a sum of $m$ terms. For other 
$l$, the ${\cal P}^{(l)}$'s vanish as there is $K\times m$ steps 
needed to connect $\sqrt[l]{\rho_{j_1}}$ and 
$\sqrt[l]{\rho_{j_l}}$ with $u_p^{\parallel}$. 
 
In the extreme case where all $N$ eigenvectors are permuted, 
only ${\cal P}_{\rho_{j_1} \ldots \rho_{j_N}}^{(N)}$ may be 
nonvanishing. Here, $K=1$ and $\det u_p^{\parallel} = 
\det U^{\parallel} =+1$ since $U^{\parallel} \in$ SU(N). 
It follows that 
\begin{equation} 
{\cal P}_{\rho_{j_1}\ldots\rho_{j_l}}^{(l)} = (-1)^{N-1} 
f_{\rho_{j_1}\ldots\rho_{j_N}}^{(N)} (\lambda_1,\ldots,\lambda_N) , 
\label{eq:permutationN} 
\end{equation} 
where each $f^{(N)}$ is determined by the sequence of $\rho$'s. 
These $f$'s have some interesting properties. First, it can be 
seen that 
\begin{equation} 
f_{\rho_{1}\ldots\rho_{N}}^{(N)} = 1, \ \forall N, 
\end{equation} 
as a consequence of the normalization condition described in Sec. 
II.C. Thus, there exist at least one well-defined off-diagonal 
mixed state phase for $U^{\parallel} = u_p^{\parallel}$, independent 
of the rank of the $\rho$'s. Secondly, we have 
\begin{equation} 
f_{\rho_{j_1}\ldots\rho_{j_N}}^{(N)} \geq 0, \ \forall 
j_1, \ldots ,j_N , 
\label{eq:maxf}
\end{equation} 
as each $f^{(N)}$ always can be written as a sum of positive 
functions of the $\lambda$'s. This implies that the off-diagonal 
mixed state phases for such unitarities are completely determined 
by the dimension of the Hilbert space ${\cal H}$. Indeed, for 
sequences where $f^{(N)} \neq 0$ we have 
\begin{eqnarray} 
\gamma^{(N)} & = & -1, \ {\textrm{if dim(${\cal H}$) even,}} 
\nonumber \\ 
\gamma^{(N)} & = & +1, \ {\textrm{if dim(${\cal H}$) odd.}} 
\end{eqnarray} 
 
Let us now turn our attention to partial permutations characterized 
by $m\neq N$, where we can use the following algorithm to determine 
$f_{\rho_{j_1}\ldots\rho_{j_l}}^{(l)}$. As in the $m=N$ case, 
each $f_{\rho_{j_1}\ldots\rho_{j_l}}^{(l)}(\lambda_1,\ldots,\lambda_N)$ 
decomposes into a sum of terms determined by the sequence of 
$\rho$'s. Explicitly, we can write $f_{\rho_{j_1} \ldots 
\rho_{j_l}}^{(l)} (\lambda_1,\ldots,\lambda_N)$ as a sum 
of $m$ terms
\begin{equation} 
\label{eq:fsum} 
f_{\rho_{j_1}\ldots\rho_{j_l}}^{(l)} (\lambda_1,\ldots,\lambda_N) = 
\sum_{i=1}^{m} A_i(\lambda_1,\ldots,\lambda_N), 
\end{equation} 
where $A_q$ is of the form $\sqrt[l]{\lambda_{j_1}^{a_1} \ldots 
\lambda_{j_l}^{a_l}} \geq 0$ with $a_i$ integers ranging from $0$ 
to $l$. Since we are not interested in the phase contributions from
$u_p^{\parallel}$ that are already included in the factor $\big[
(-1)^{m-1} \det u_p^{\parallel} \big]^K$, we replace $u_p^{\parallel}$ 
with the operator 
\begin{equation} 
U_g^{(m)} \equiv 
\ket{\psi_m}\bra{\psi_1} + \ket{\psi_1}\bra{\psi_2} + 
\ldots \ket{\psi_{m-1}}\bra{\psi_m} , 
\label{eq:permop} 
\end{equation} 
being unitary on the permuted subspace. Thereafter, we compute 
$f_{\rho_{j_1}\ldots\rho_{j_l}}^{(l)} (\lambda_1,\ldots,\lambda_N) = 
\Tr \big( W^{j_1} \ldots W^{j_l} \big)$ with $W^{j_k} \equiv 
U_g^{(m)} \sqrt[l]{\rho_{j_k}}$. Applying ordinary matrix
multiplication rules and noting that only one entry in each row and
column of $W^{j_k}$ is nonvanishing, $A_i$ is in index notation given
by
\begin{eqnarray} 
\label{eq:Asplit} 
A_i & = & W^{j_1}_{i,(i+1)\mod m} W^{j_2}_{(i+1)\mod m,(i+2)\mod m} 
\ldots 
\nonumber \\ 
 & & \times W^{j_l}_{(i+l-1)\mod m,i}. 
\end{eqnarray} 
Each eigenvalue $\lambda_k$ of $\rho_1$ appears exactly once in each 
$W^{j_k}$, so our aim is to describe the correspondence between the 
components $W^{j_k}_{x,y}$ and $\lambda_k$ by appropriate index 
transformations. 
 
To revert the off-diagonal matrices $W^{j_l}$ to diagonal form we 
apply $\big( U_g^{(m)}\big)^\dagger$ to each $W^{j_k}$, thus $W^{j_k} 
\rightarrow \big( \sqrt[l]{\rho_{j_k}} \big) = \big(U_g^{(m)} 
\big)^\dagger W^{j_k}$. This transforms the indexes according to 
\begin{eqnarray} 
\label{eq:rule1} 
x & \rightarrow & x^\prime = (x+1) \mod m ,  
\nonumber \\ 
y & \rightarrow & y^\prime = y . 
\end{eqnarray} 
leading to 
\begin{equation}
W^{j_k}_{x,y} = \big( \sqrt[l]{\rho_{j_k}} \big)_{(x+1) \mod m,y} 
\end{equation} 
in terms of components. Consequently we obtain 
\begin{eqnarray} 
\label{eq:Asplitorig} 
A_i & = & 
\big( \sqrt[l]{\rho_{j_1}} \big)_{(i+1)\mod m,(i+1)\mod m} \ldots  
\nonumber \\ 
 & & \times 
\big( \sqrt[l]{\rho_{j_{l-1}}} \big)_{(i+l-1)\mod m,(i+l-1)\mod m} 
\big( \sqrt[l]{\rho_{j_l}} \big)_{i,i}. 
\nonumber \\ 
\end{eqnarray} 
The transformation back to the ``unpermuted'' $\rho_1$ can be 
achieved by applying $U_g$ described in Eq. (\ref{eq:Uperm}) 
accomplishing $\ket{\psi_n} = \big( U_g \big)^{n-1} \ket{\psi_1}, \ 
n=1,\ldots ,N$, thus $\sqrt[l]{\rho_{j_k}} \rightarrow 
\sqrt[l]{\rho_1} = (U_g^{\dagger})^{j_k-1} \sqrt[l]{\rho_{j_k}} 
(U_g)^{j_k-1}$. Since $j_k - 1$ steps are needed to convert 
$\sqrt[l]{\rho_{j_k}}$ to $\sqrt[l]{\rho_{1}}$ we have to carry 
out the index transformation 
\begin{equation} 
\label{eq:rule2} 
x \rightarrow x^\prime = \big( x-(j_k-1)\big) \mod N 
\end{equation} 
so that $\big( \sqrt[l]{\rho_{j_k}} \big)_{x,x} \rightarrow \big( 
\sqrt[l]{\rho_{1}} \big)_{x',x'}$. Since $\rho_1$ is diagonal with 
eigenvalues $\lambda_k$ in ascending order in $k$, the index 
$x^\prime$ denotes the wanted eigenvalue $\lambda_{x'}$. 
 
This algorithm traces the locations of the eigenvalues $\lambda_k$ 
from $W^{n}$ back to $\sqrt[l]{\rho_1}$ when the unitary 
transformations along the path starting from $W^{n}$ are applied: 
\begin{equation} 
\begin{CD} 
\sqrt[l]{\rho_1} @>{\left(U_g\right)^{n-1} \ldots 
\left(U_g^\dagger\right)^{n-1}}>>  \sqrt[l]{\rho_n}\\ 
 @A{\left(U_g^\dagger \right)^{n-1} \ldots \left( U_g \right)^{n-1}}AA
 @VV{U_g^{(m)}}V\\
\sqrt[l]{\rho_n} @<<{\big(U_g^{(m)}\big)^{\dagger}}< W^{n} 
\end{CD} 
\end{equation} 
This analysis makes it possible to calculate the factor
$f_{\rho_{j_1}\ldots\rho_{j_l}}^{(l)} (\lambda_1,\ldots,\lambda_N)$
more efficiently than performing a multiplication of the $l$ matrices
involved. 
 
Let us revisit the qubit $(N=2)$ case using the above general 
theory. If $m=0$, both $\gamma_{\rho_1}^{(1)}$ and 
$\gamma_{\rho_2}^{(1)}$ exist. Moreover, we have 
\begin{eqnarray} 
{\cal D}_{\rho_{1} \rho_{2}}^{(2)} = \sqrt{\lambda_1\lambda_2} 
\big[ \big( U_{11}^{\parallel} \big)^2 + 
\big( U_{22}^{\parallel} \big)^2 \big] , 
\end{eqnarray} 
which is consistent with Eq. (\ref{eq:offqubit}) for $\eta=1$. 
In the permutation case $m=2$, we may use $(-1)^{N-1} 
\det U^{\parallel} = -1$ and, from Eq. (\ref{eq:maxf}), 
$f_{\rho_1 \rho_2}^{(2)} = 1$ for $N=2$, and obtain 
\begin{eqnarray} 
{\cal P}_{\rho_{1} \rho_{2}}^{(2)} = -1, 
\end{eqnarray} 
in agreement with Eq. (\ref{eq:offqubit}) for $\eta=0$. 
 
As a further illustration, let us work out the $N=3$ case 
in detail. For $m=0$, all the $\gamma^{(1)}$'s are well-defined. 
The dependence upon the rank of the density operator is visible 
for higher $l$, namely 
\begin{eqnarray} 
{\cal D}_{\rho_{1} \rho_{2}}^{(2)} & = & \sqrt{\lambda_1 \lambda_3} 
\big( U_{11}^{\parallel} \big)^2 + \sqrt{\lambda_1 \lambda_2} 
\big( U_{22}^{\parallel} \big)^2 
\nonumber \\ 
 & & + \sqrt{\lambda_2 \lambda_3} 
\big( U_{33}^{\parallel} \big)^2 , 
\nonumber \\ 
{\cal D}_{\rho_{1} \rho_{2} \rho_{3}}^{(3)} & = & 
{\cal D}_{\rho_{1} \rho_{3} \rho_{2}}^{(3)}
\nonumber \\ 
 & = & \sqrt[3]{\lambda_1 \lambda_2 \lambda_3} 
\big[ \big( U_{11}^{\parallel} \big)^3 + 
\big( U_{22}^{\parallel} \big)^3 + 
\big( U_{33}^{\parallel} \big)^3 \big] 
\nonumber \\ 
\end{eqnarray} 
with ${\cal D}_{\rho_{2} \rho_{3}}^{(2)}$ and 
${\cal D}_{\rho_{3} \rho_{1}}^{(2)}$ obtained by permutations 
of the $\lambda$'s. In the $m=2$ case, $\ket{\psi_1} \rightarrow 
\ket{\psi_2} \rightarrow \ket{\psi_1}$ while $\ket{\psi_3}$ 
undergoes cyclic evolution. Explicitly we have 
\begin{eqnarray} 
{\cal D}_{\rho_{1}}^{(2)} & = & \lambda_3  U_{33}^{\parallel} , 
\nonumber \\ 
{\cal D}_{\rho_{1} \rho_{2}}^{(2)} & = & 
\sqrt{\lambda_2 \lambda_3}  \big( U_{33}^{\parallel} \big)^2 , 
\nonumber \\ 
{\cal D}_{\rho_{1} \rho_{2} \rho_{3}}^{(3)} & = & 
{\cal D}_{\rho_{1} \rho_{3} \rho_{2}}^{(3)} = 
\sqrt[3]{\lambda_1 \lambda_2 \lambda_3} 
\big( U_{33}^{\parallel} \big)^3 .  
\end{eqnarray} 
${\cal P}_{\rho_{1} \rho_{2}}^{(2)}$ can be calculated via 
the algorithm above as follows. Write $f^{(2)}_{\rho_1 \rho_2} = 
A_1 + A_2$ with $A_1 = W^{1}_{1,2} W^{2}_{2,1}$ and 
$A_2 = W^{1}_{2,1} W^{2}_{1,2}$. Application of the 
rule in Eq. (\ref{eq:rule1}) yields 
\begin{eqnarray}
A_1 & = & \big(\sqrt{\rho_1}\big)_{2,2} 
\big(\sqrt{\rho_2}\big)_{1,1}, 
\nonumber \\ 
A_2 & = & \big(\sqrt{\rho_1}\big)_{1,1} 
\big(\sqrt{\rho_2}\big)_{2,2}
\end{eqnarray} 
and after using the rule in Eq. (\ref{eq:rule2}) we obtain 
\begin{eqnarray} 
A_1 & =& \big(\sqrt{\rho_1}\big)_{2,2} 
\big(\sqrt{\rho_1}\big)_{3,3}= \sqrt{\lambda_2 \lambda_3} ,  
\nonumber \\ 
A_2 &=& \big(\sqrt{\rho_1}\big)_{1,1}\big( 
\sqrt{\rho_1}\big)_{1,1} = \lambda_1 .  
\end{eqnarray} 
Thus, we obtain 
\begin{equation}
{\cal P}_{\rho_{1} \rho_{2}}^{(2)} = 
U_{12}^{\parallel} U_{21}^{\parallel} 
\left( \lambda_{1} + \sqrt{\lambda_2 \lambda_3} \, \right) , 
\end{equation} 
where we have used $(-1)^{m-1} \det u_{p}^{\parallel} = 
U_{12}^{\parallel} U_{21}^{\parallel}$. The remaining 
${\cal D}_{\rho_{2}}^{(1)}$, ${\cal D}_{\rho_{3}}^{(1)}$, 
${\cal D}_{\rho_{2} \rho_{3}}^{(2)}$, ${\cal D}_{\rho_{3} 
\rho_{1}}^{(2)}$, ${\cal P}_{\rho_{2} \rho_{3}}^{(2)}$, 
${\cal P}_{\rho_{3} \rho_{1}}^{(2)}$ are given by appropriate 
permutations of the $\lambda$'s. For $m=3$ (full permutation) 
the only possible contributions are 
\begin{eqnarray} 
{\cal P}_{\rho_{1} \rho_{2} \rho_{3}}^{(3)} & = & 1 , 
\nonumber \\ 
{\cal P}_{\rho_{1} \rho_{3} \rho_{2}}^{(3)} & = & 
3 \sqrt[3]{\lambda_1 \lambda_2 \lambda_3}  , 
\end{eqnarray} 
where we have used $(-1)^{N-1} \det U^{\parallel} = +1$ and, again 
from Eq. (\ref{eq:maxf}), $f_{\rho_1 \rho_2 \rho_3}^{(3)} = 1$ for 
$N=3$. The expression for ${\cal P}_{\rho_{1} \rho_{3} \rho_{2}}^{(3)}$ 
follows from $A_1 = A_2 = A_3 = \sqrt[3]{\lambda_1 \lambda_2 \lambda_3}$ 
and requires full rank to be nonvanishing.  

As an illustrative higher dimensional example we consider the 
case where $N=5$, $l=4$ and a partial permutation specified 
by $m=2$. The diagonal part can be calculated to
\begin{eqnarray} 
{\cal D}^{(4)}_{\rho_1 \rho_4 \rho_5 \rho_3} & = & U_{33}^{\parallel} 
\sqrt[4]{\lambda_1 \lambda_3 \lambda_4 \lambda_5} + U_{44}^{\parallel} 
\sqrt[4]{\lambda_1 \lambda_2 \lambda_4 \lambda_5} 
\nonumber \\ 
 & & + U_{55}^{\parallel}  
\sqrt[4]{\lambda_1 \lambda_2 \lambda_3 \lambda_5} 
\end{eqnarray} 
and the permutation part is given by
\begin{equation} 
{\cal P}^{(4)}_{\rho_1 \rho_4 \rho_5 \rho_3} = \big[ (-1)^{1} \det
u_p^{\parallel} \big]^2 f_{\rho_{1} \rho_4 \rho_5 \rho_3}^{(4)},
\end{equation} 
where $f^{4}_{\rho_1\rho_4\rho_5\rho_3} =
A_1(\lambda_1,\ldots,\lambda_5) + A_2(\lambda_1,\ldots,\lambda_5)$. 
For a calculation of $f^{(4)}_{\rho_1\rho_4\rho_5\rho_3}$ we use 
the algorithm described above. From Eq.~(\ref{eq:Asplit}) we know 
that
\begin{eqnarray} 
A_1 & = & W^1_{1,2} W^4_{2,1} W^5_{1,2} W^3_{2,1}, 
\nonumber \\ 
A_2 & = & W^1_{2,1} W^4_{1,2} W^5_{2,1} W^3_{1,2}. 
\end{eqnarray} 
Applying rule (\ref{eq:rule1}) to the indexes we obtain 
\begin{eqnarray} 
A_1 & = & \big(\sqrt[4]{\rho_1}\big)_{2,2}
\big(\sqrt[4]{\rho_4}\big)_{1,1} \big(\sqrt[4]{\rho_5}\big)_{2,2}
\big(\sqrt[4]{\rho_3}\big)_{1,1}, 
\nonumber \\ 
A_2 & = &
\big(\sqrt[4]{\rho_1}\big)_{1,1} \big(\sqrt[4]{\rho_4}\big)_{2,2}
\big(\sqrt[4]{\rho_5}\big)_{1,1} \big(\sqrt[4]{\rho_3}\big)_{2,2} . 
\end{eqnarray} 
After a transformation according to the rule (\ref{eq:rule2}) 
we have 
\begin{eqnarray} 
A_1 & = & \big(\sqrt[4]{\rho_1}\big)_{2,2} \big(
\sqrt[4]{\rho_1}\big)^2_{3,3} \big( \sqrt[4]{\rho_1}\big)_{4,4} =
\sqrt[4]{\lambda_2 \lambda_3^2 \lambda_4}, 
\nonumber \\ 
A_2 & = &
\big(\sqrt[4]{\rho_1}\big)_{1,1} \big( \sqrt[4]{\rho_1}\big)_{4,4}
\big( \sqrt[4]{\rho_1}\big)_{2,2} \big( \sqrt[4]{\rho_1}\big)_{5,5}
\nonumber \\ 
 & = & \sqrt[4]{\lambda_1 \lambda_4 \lambda_2 \lambda_5}
\end{eqnarray} 
and consequently 
\begin{equation} 
f^{(4)}_{\rho_1\rho_4\rho_5\rho_3} = A_1 + A_2 = \sqrt[4]{\lambda_2 
\lambda_3^2 \lambda_4} + \sqrt[4]{\lambda_1 \lambda_2 \lambda_4 
\lambda_5}. 
\end{equation} 
${\cal P}^{(4)}_{\rho_1 \rho_4 \rho_5 \rho_3}$ can now be written as 
\begin{eqnarray} 
{\cal P}^{(4)}_{\rho_1 \rho_4 \rho_5 \rho_3} = \big(U_{12}^{\parallel}
U_{21}^{\parallel}\big)^2 \left( \sqrt[4]{\lambda_2 \lambda_3^2 \lambda_4} +
\sqrt[4]{\lambda_1 \lambda_2 \lambda_4 \lambda_5} \right),
\nonumber \\ 
\end{eqnarray} 
using $\det u_{p}^{\parallel}= -U_{12}^{\parallel}U_{21}^{\parallel}.$ 
  
\section{Experimental procedure} 
When we consider possible experimental realizations of the 
off-diagonal mixed state phases we immediately encounter a problem: 
how do we experimentally implement the $l$th root of density operators? 
Fortunately, this may be resolved in the $l=2$ case in the sense of 
purification, i.e., by adding an ancilla system in a certain way. 
Here, we demonstrate this in general and propose a physical 
scenario for the qubit case in terms of polarization-entangled 
two-photon interferometry. 

We first show how to realize the $l=1$ and $l=2$ phases via 
purification. For an $N$ dimensional Hilbert space ${\cal H}$, 
consider the nondegenerate density operator
\begin{equation} 
\rho_1 = 
\sum_{k=1}^N \lambda_k \ket{\psi_{k}} \bra{\psi_{k}} . 
\end{equation}
A purification of this $\rho_1$ is any pure state $\Psi_1$ obtained 
by adding an ancilla system $a$ to the considered system $s$ such 
that $\rho_1 = \Tr_a \ket{\Psi_1} \bra{\Psi_1}$. Thus, we may write 
\begin{equation} 
\ket{\Psi_1} = 
\sum_{k=1}^N \sqrt{\lambda_k} \ket{\psi_{k}} \otimes \ket{\varphi_{k}} , 
\label{eq:pur1}
\end{equation}
where $\{ \ket{\varphi_{k}} \}$ is an orthonormal set of vectors in 
the ancilla Hilbert space ${\cal H}_a$. Consequently, any orthogonal 
density operator $\rho_{n} = \big( U_g \big)^{n-1} \rho_1 \big( 
U_g^{\dagger} \big)^{n-1}$ has a purification of the form  
\begin{equation} 
\ket{\Psi_n} = \big( U_g \big)^{n-1} \otimes \widetilde{U}_a \ket{\Psi_1} 
\end{equation} 
for any unitarity $\widetilde{U}_a$ acting on ${\cal H}_a$. In the 
following, we assume $\dim {\cal H}_a = N$ and put $\ket{\varphi_k} = 
\ket{\psi_k}$. 

Let $U_s \otimes U_a \ket{\Psi_1}$ and $V_s \otimes V_a \ket{\Psi_1}$
be two Hilbert space representatives of a pair of purifications of
$U_s \rho_1 U_s^{\dagger}$ and $V_s \rho_1 V_s^{\dagger}$. The
coincidence interference pattern obtained in superposition is
determined by the interference profile
\begin{eqnarray} 
{\cal I} & \propto & \Big| U_s \otimes U_a \ket{\Psi_1} + V_s 
\otimes V_a \ket{\Psi_1} \Big|^{2} 
\nonumber \\ 
 & = & 2 + 2 {\textrm{Re}} \big[ \Tr \big( U_s^{\dagger} V_s \otimes 
U_a^{\dagger} V_a \ket{\Psi_1} \bra{\Psi_1} \big) \big] . 
\end{eqnarray}
By choosing $U_s = e^{i\chi} \big( U_g \big)^{j_1-1}$, $V_s = 
U^{\parallel} \big( U_g \big)^{j_1-1}$, and $U_a = V_a = I$, 
we obtain the $l=1$ phase factors $\gamma_{\rho_{j_1}}^{(1)}$ 
by variation of the U(1) phase $\chi$ since  
\begin{eqnarray} 
& & \Phi \big[ \Tr \big( U_s^{\dagger} V_s \otimes U_a^{\dagger} V_a 
\ket{\Psi_1} \bra{\Psi_1} \big) \big] 
\nonumber \\ 
 & & =  e^{-i\chi} \Phi \big[ \Tr \big( \big( U_g^{\dagger} 
\big)^{j_1-1} U^{\parallel}\big( U_g \big)^{j_1-1} \otimes 
I \ket{\Psi_1} \bra{\Psi_1} \big)\big]
\nonumber \\ 
 & & = e^{-i\chi} \Phi \big[ \Tr \big( U^{\parallel} 
\rho_{j_1} \big) \big] , 
\end{eqnarray}
where we have used that $\Tr_a \big[ \big( U_g \big)^{j_1-1} 
\ket{\Psi_1} \bra{\Psi_1} \big( U_g^{\dagger} \big)^{j_1-1} \big]$ $= 
\rho_{j_1}$. Similarly, the $l=2$ phase factors 
$\gamma_{\rho_{j_1}\rho_{j_2}}^{(2)}$ are obtained 
by letting $U_s = e^{i\chi} \big( U_g \big)^{j_2-1}$, $V_s = 
U^{\parallel} \big( U_g \big)^{j_1-1}$, $U_a = \big( 
U_g \big)^{j_2-1}$, and $V_a = \big( U^{\parallel} \big)^{\textrm{T}} 
\big( U_g \big)^{j_1-1}$, T being transpose with respect to the 
ancilla basis $\{ \ket{\psi_{k}} \}$, since   
\begin{eqnarray}
 & & \Phi \big[ \Tr \big( U_s^{\dagger} V_s \otimes 
U_a^{\dagger} V_a \ket{\Psi_1} \bra{\Psi_1} \big) \big] 
\nonumber \\ 
 & & = e^{-i \chi} \Phi \big[ \Tr \big( \big( U_g^{\dagger} \big)^{j_2-1} 
U^{\parallel} \big( U_g \big)^{j_1-1} 
\nonumber \\ 
 & & \otimes \big( U_g^{\dagger} 
\big)^{j_2-1} \big( U^{\parallel} \big)^{\textrm{T}} 
\big( U_g \big)^{j_1-1} \ket{\Psi_1} \bra{\Psi_1} \big) \big]  
\nonumber \\ 
 & & = e^{-i \chi} \Phi \big[ \Tr \big( U^{\parallel} 
\sqrt{\rho_{j_1}} U^{\parallel} \sqrt{\rho_{j_2}} \big) \big] , 
\end{eqnarray}
where the last equality may be obtained by explicit use of 
$\ket{\Psi_1}$ in Eq. (\ref{eq:pur1}) with $\ket{\varphi_k} = 
\ket{\psi_k}, \ \forall k$. 

\begin{figure}[ht!] 
\begin{center} 
\includegraphics[width=8 cm]{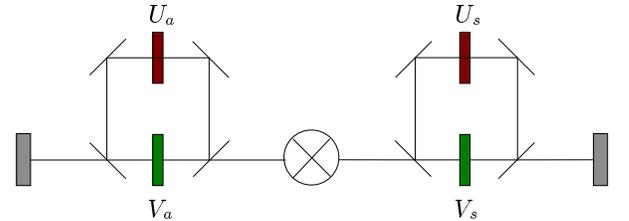} 
\end{center} 
\caption{Franson setup for polarization-entangled photon pairs. 
In the longer arms, the system and ancilla photons are exposed to the 
polarization affecting unitarities $U_s$ and $U_a$, respectively, and 
similarly $V_s$ and $V_a$ in the shorter arms.} 
\label{fig:franson}
\end{figure}

Let us discuss a physical purification scenario for the $l=1$ and
$l=2$ phases in the qubit case. Consider the two-photon Franson-type
\cite{franson89} setup in Fig. \ref{fig:franson}. A source that in the
horizontal-vertical $(h-v)$ basis produces polarization-entangled
photon states of the form 
\begin{eqnarray} 
\ket{\Psi_1} & = & \sqrt{\frac{1}{2}(1+r)} \ket{h} 
\otimes \ket{h}
\nonumber \\ 
 & & + \sqrt{\frac{1}{2}(1-r)} \ket{v} 
\otimes \ket{v}  
\end{eqnarray} 
has been demonstrated in Ref. \cite{kwiat99}. Considered as 
subsystems both photons are in a mixed linear polarization 
state $\rho_1$ with polarization degree $r$. The desired 
superposition of $U_s \otimes U_a \ket{\Psi_1}$ and 
$V_s \otimes V_a \ket{\Psi_1}$ is obtained by requiring 
sufficiently short coincidence window so that detection 
occurs only when the photons both either took the shorter 
path or the longer path \cite{hessmo00}. A purification of 
the orthogonal density operator $\rho_2 = U_g \rho_1
U_g^{\dagger}$ may be achieved by flipping the polarizations 
of the photons, yielding
\begin{eqnarray} 
\ket{\Psi_2} = U_g \otimes U_g \ket{\Psi_1} & = & 
\sqrt{\frac{1}{2}(1+r)} \ket{v} \otimes \ket{v} 
\nonumber \\ 
 & & + \sqrt{\frac{1}{2}(1-r)} \ket{h} \otimes \ket{h} .   
\end{eqnarray} 

\begin{figure}[ht!] 
\begin{center} 
\includegraphics[width=8 cm]{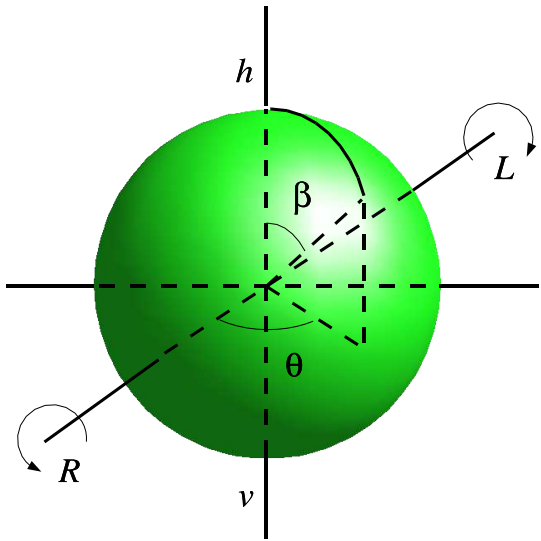} 
\end{center} 
\caption{Effect of the unitarity $U(\beta,\theta)$ on the Poincar\'{e} 
sphere. The horizontal polarization state $h$ at the north pole is 
taken into a new polarization state at spherical polar angles 
$(\beta,\theta)$.} 
\label{fig:poincare}
\end{figure}

To demonstrate the $l=1$ and $l=2$ geometric phases in this scenario, 
it is sufficient to consider unitarities that rotate linear polarization 
states along great circles an angle $\beta$ on the Poincar\'{e} sphere, 
see Fig. \ref{fig:poincare}. This amounts to 
\begin{eqnarray} 
U (\beta,\theta) & = & 
\exp \Big( -i\frac{\beta}{2} \Big[ \cos \theta \big( \ket{h}\bra{v} + 
\ket{v}\bra{h} \big)  
\nonumber \\ 
 & & + \sin \theta \big( -i\ket{h}\bra{v}  
 + i\ket{v} \bra{h} \big) 
\Big] \Big) , 
\label{eq:polarrotation} 
\end{eqnarray} 
which fulfills the parallel transport conditions in Eqs. (\ref{eq:pcj}) 
and (\ref{eq:pch}) with respect to the $h-v$ basis. In practice, 
$U (\beta,\theta)$ may be implemented by appropriate $\lambda-$plates, 
the thickness and orientation of which correspond to the parameters 
$\beta$ and $\theta$, respectively. For example, $U_g=U(\pi ,\pi /2)$ 
acting on the linear polarization states as a polarization flip and 
thus connects $\rho_1$ and $\rho_2$, is achieved by a $\lambda /2$ 
plate with half axis making an angle $45^{\circ}$ to the vertical ($v$) 
direction.  Furthermore, $\theta =0$ and $\beta = \pi /2$, corresponding 
to a $\lambda /4$ plate oriented along the vertical direction, takes
$h$ and $v$ into the right ($R$) and left ($L$) circular polarization 
states, respectively.

The phase factors $\gamma_{\rho_n}^{(1)}$, $n=1,2$, are obtained from
the coincidence intensity by choosing $U_s = e^{i\chi}
\big( U_g \big)^{n-1}$, $V_s = U(\beta,\theta) \big( U_g \big)^{n-1}$, 
and $U_a = V_a = I$. Explicit calculation for $U(\beta,\theta)$ in 
Eq. (\ref{eq:polarrotation}) yields $\Tr \big( \rho_1 
U(\beta,\theta) \big) = \Tr \big( \rho_2  U(\beta,\theta) 
\big) = \cos (\beta /2)$, which entails that 
$\gamma_{\rho_1}^{(1)}$ and $\gamma_{\rho_2}^{(1)}$ are real-valued 
and changes sign at $\beta = (2j+1)\pi$, $j$ integer, 
corresponding to a sequence of phase jumps of $\pi$.
Furthermore, the choice $U_s = e^{i\chi} U_g$, $V_s = U(\beta,\theta)$, 
$U_a = U_g$, and $V_a = U^{{\textrm{T}}} (\beta,\theta) = 
U(\beta,-\theta)$, yields $\gamma_{\rho_1\rho_2}^{(2)}$ and we 
may compute the expected output as $\Tr \big( \sqrt{\rho_1} 
U(\beta,\theta) \sqrt{\rho_2} U(\beta,\theta) \big) = 
\sqrt{1-r^2} \cos^{2} \big(\beta /2\big) - \sin^{2} \big(\beta /2\big)$, 
which is independent of $\theta$ and can be positive and negative for 
$r\neq 1$ depending upon $\beta$. $\gamma_{\rho_1\rho_2}^{(2)}$ changes 
sign at $\beta = 2\pi j + 2 \arctan \sqrt[4]{1-r^2}$. Note that 
$0\leq \arctan \sqrt[4]{1-r^2} \leq \pi /4$, modulus $\pi$, which 
assures that the $l=1$ and $l=2$ phases never become indeterminate 
for the same $\beta$ value, and thus provide a complete experimental 
phase characterization of the qubit case in the sense of purification. 

\section{Projection phase} 
Since the definition of the off-diagonal geometric mixed state phase 
claims to be reducible to the pure state off-diagonal geometric phase, 
the question arises if there is a connection to the experimental 
verification of the latter performed by Hasegawa {\it et al.} 
\cite{hasegawa01,hasegawa02}. This has to be answered in the 
negative, since in this experiment the evolution of the orthogonal
state is implemented as a projection operator, which is by definition
equivalent to a pure state. On the other hand, by taking the impurity
of the input state into account, the shift in the interference pattern
in the Hasegawa {\it et al.} setup, given by the additional phase factor 
\begin{equation} 
\label{eq:projphase} 
\gamma_{\rho P} \equiv \Phi \big[ \Tr \big(U \rho U P \big) \big], 
\end{equation} 
could be used as a definition of the off-diagonal geometric mixed
state phase, if the unitarity $U$ describing the evolution inside the
interferometer is parallel transporting the eigenvectors of the
nondegenerate $\rho$. Here, $P$ should project onto the eigenstate 
that corresponds to the smallest eigenvalue of $\rho$. Parallel 
transport is for example fulfilled in the Hasegawa {\it et al.} 
experiment if the incident spinor is polarized in a plane
perpendicular to the direction of the magnetic field. 
$\gamma_{\rho P}$ is gauge invariant under a U(1) transformation 
of each of the basis vectors and it reduces to the corresponding 
off-diagonal geometric phase factor of Ref. \cite{manini00} in 
the pure state limit when $U=U^{\parallel}$.
 
In the two dimensional case relevant for the Hasegawa {\it et al.} 
experiment with input $\rho = \lambda_1 \ket{\psi_1} \bra{\psi_1} +  
\lambda_2 \ket{\psi_2} \bra{\psi_2}$, $\lambda_1 > \lambda_2$, 
we can write Eq. (\ref{eq:projphase}) as 
\begin{equation} 
\label{eq:projphase2} 
\Tr[U \rho U P] = \lambda_1 (-1 + \eta^2 ) + 
\lambda_2 \eta^2 e^{-2 i \alpha} .  
\end{equation} 
Here, $U\in$ SU(2) with the diagonal matrix elements $U_{11} = 
U_{22}^{\ast} = \eta e^{i \alpha}$ is not necessarily fulfilling 
the parallel transport condition with respect to $\{ \ket{\psi_1}, 
\ket{\psi_2} \}$. In the pure state limit $\lambda_1 = 1,\ 
\lambda_2 = 0$ the off-diagonal phase is always $\pi$ since 
$\Tr[U\rho UP]_{\lambda_1=1} = -1 + \eta^2$ is real and negative, 
irrespective of whether $U$ parallel transports $\ket{\psi_1},
\ket{\psi_2}$ or not. For a mixed input state $\rho$ the 
$\lambda_2$-term does not vanish and we obtain additional 
geometric and/or dynamical phase contributions. These can be 
considered to originate in the subjacent geometry only if $U$ 
is a parallel transporting unitarity, but not for arbitrary $U$.
  
\begin{figure}[ht!] 
\centering 
\includegraphics[width=60mm]{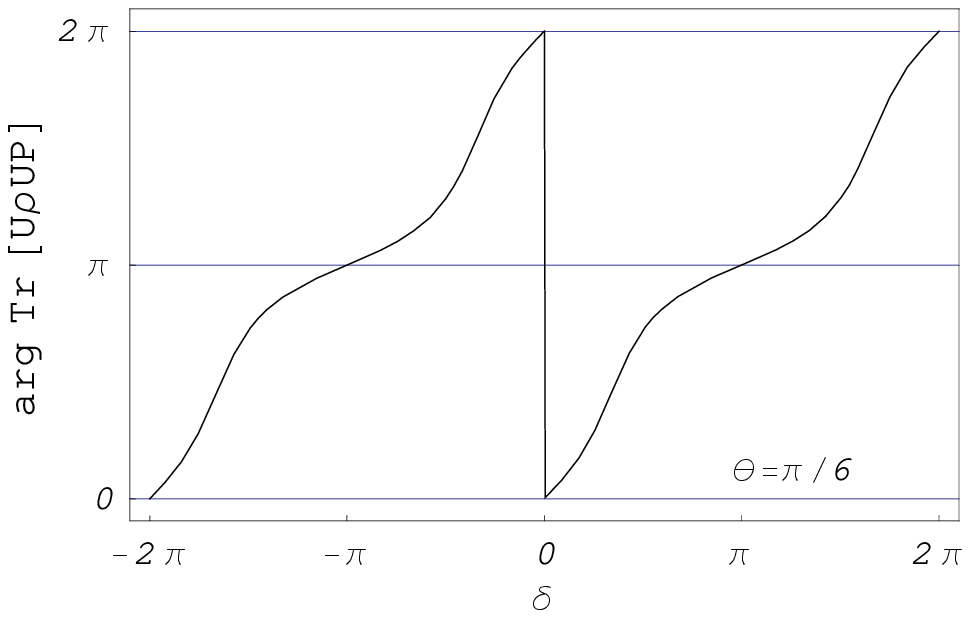} 
\hspace{20pt} 
\includegraphics[width=60mm]{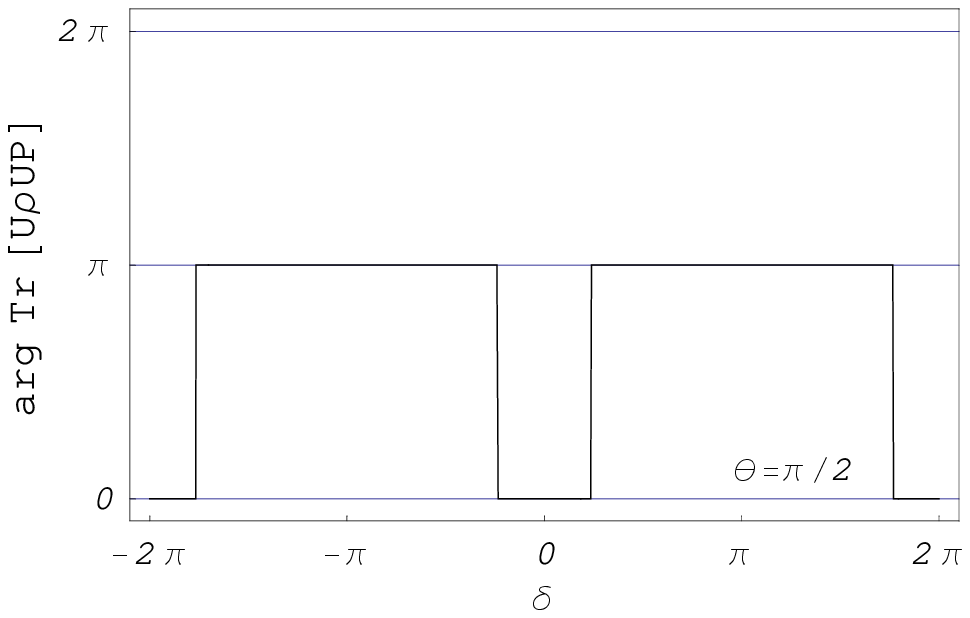} 
\caption{The projection off-diagonal mixed state geometric phase 
$\arg\Tr[U \rho U P]$ in radians modulus $2\pi$ for the Hasegawa 
setup with $\lambda_1 = 0.87,\lambda_2=0.13$ and $\theta=\pi/6$ or
$\theta=\pi/2$, respectively . $\delta$ is the precession angle in 
radians of the incident spinor about the direction of the magnetic 
field.}  
\label{fig:projphase}
\end{figure}

To show the consistency with the experiment performed by Hasegawa
{\it et al.} we calculate the phase $\phi_{\rho P} = 
\arg\Tr[U\rho U P]$. In the left panel of Fig. \ref{fig:projphase} 
we show $\phi_{\rho P}$ for a mixed input state with $\lambda_1 =
0.87, \lambda_2 = 0.13$, in accordance with the experimental degree 
of polarization in Refs. \cite{hasegawa01,hasegawa02}, and the spin
polarization angle $\theta = \pi/6$ relative to the magnetic field in
the upper arm of the interferometer (see Fig. 2 of
Ref. \cite{hasegawa02}). The calculated curve matches with the
experimental and theoretical results presented in Fig. 5(d) of
Ref. \cite{hasegawa02}. Note that in this case $U$ is not parallel
transporting the incident spinor.

Another interesting fact is that due to the impurity of the input 
state we expect phase jumps for $\theta=\pi/2$ for $\delta=2\arccos 
\sqrt{\lambda_1}$ and $\delta = 2\pi - 2\arccos \sqrt{\lambda_1}$, 
see right panel of Fig. \ref{fig:projphase}, where $\delta$ is the 
precession angle of the incident spinor about the direction of the 
magnetic field. Here, we have a parallel transporting $U=U^\parallel$, 
thus these jumps have their origin in the subjacent geometry of state 
space. 

The projection off-diagonal geometric phase factor $\gamma_{\rho P}$
is invariant under phase transformations of the eigenvectors of $\rho$,
it reduces to the corresponding $l=2$ phase factor of Ref. 
\cite{manini00} in the pure state limit, and it has the advantage
that it can be observed in single particle experiments. The drawbacks
are that it is less symmetric than that in Eq. (\ref{eq:mixedl2}), one
cannot easily state a generalization like in
Eq. (\ref{eq:genoffdiag}), and vice versa it cannot be regarded an 
off-diagonal generalization of the mixed state geometric phase in
Ref. \cite{sjoqvist00} as $\rho$ and $P$ are not unitarily
connected. These features suggest that the mixed state geometric phase
factor in Eq. (\ref{eq:mixedl2}) is to prefer over $\gamma_{\rho P}$.
 
\section{Conclusions}
Recent investigations in geometric phases in quantum systems have
led to cases where the standard definitions breaks down. On one
hand, such situations emerge for orthogonal initial and final pure
states connected unitarily, on the other, unitary evolution of a
system in a mixed state may lead to nodal points in parameter space. 
In search for a complementary geometric quantity defined in such 
cases, the off-diagonal mixed state geometric phase has been proposed 
\cite{filipp03} by the present authors as a generalization of 
the off-diagonal geometric phases for pure state put forward in 
Ref. \cite{manini00}.

Starting with a preliminary discussion about orthogonality of mixed
states and quantum parallel transport we have provided a general
treatment of the off-diagonal mixed state geometric phase comprising
unitarities that can be decomposed into a diagonal part leaving the
initial basis states unchanged and a permutation part reordering the
initial states. An algorithm has been presented to calculate the
appropriate phase factors efficiently for any dimension and for
an arbitrary number of orthogonal density operators. Furthermore, we
have discussed the projection off-diagonal geometric phase appearing
in the neutron experiment by Hasegawa {\it et al.} 
\cite{hasegawa01,hasegawa02} as an alternative definition 
of off-diagonal mixed state phase. 

In the qubit case the off-diagonal mixed state phase can be fully
qualified both from the theoretical and from the experimental point of
view. But it has to be mentioned that the measurement seems to require
control and measurement of one or more ancilla systems although the
off-diagonal mixed state phases are properties of the system alone,
since the constituting set of density operators pertains solely to the
system. Explicitly, a Franson interferometer setup for the qubit case
has been presented illustrating the nontrivial sign change property of
the off-diagonal phase connected to the mixed state case. The apparent 
need for control over an ancilla system seems to suggest that the 
proposed concept of off-diagonal mixed state geometric phase is a 
nonlocal and/or contextual property of the unitary evolution of a 
quantum system.

\section*{Acknowledgments} 
S.F. acknowledges support from the Austrian Science Foundation, 
Project No. F1513. The work by E.S. was supported by the Swedish 
Research Council.


\begin{thebibliography}{99} 
\bibitem{pancharatnam56} S. Pancharatnam, 
Proc. Indian Acad. Sci. A {\bf 44}, 247 (1956). 
\bibitem{berry84} M.V. Berry, 
Proc. Roy. Soc. London Ser. A {\bf 392}, 45 (1984). 
\bibitem{aharonov87} Y. Aharonov and J. Anandan, 
Phys. Rev. Lett. {\bf 58}, 1593 (1987). 
\bibitem{samuel88} J. Samuel and R. Bhandari, 
Phys. Rev. Lett. {\bf 60}, 2339 (1988). 
\bibitem{uhlmann86} A. Uhlmann,
Rep. Math. Phys. {\bf 24}, 229 (1986).
\bibitem{sjoqvist00} E. Sj\"{o}qvist, A.K. Pati, A. Ekert, 
J.S. Anandan, M. Ericsson, D.K.L. Oi, and V. Vedral, 
Phys. Rev. Lett. {\bf 85}, 2845 (2000).
\bibitem{slater02} P.B. Slater, 
Lett. Math. Phys. {\bf 60}, 123 (2002); e-print math-ph/0112054;  
M. Ericsson, A.K. Pati, E. Sj\"{o}qvist, 
J. Br\"{a}nnlund, and D.K.L. Oi, 
Phys. Rev. Lett. (to be published). 
\bibitem{ericsson03} M. Ericsson, E. Sj\"{o}qvist, 
J. Br\"{a}nnlund, D.K.L. Oi, and A.K. Pati,  
Phys. Rev. A {\bf 67}, 020101(R) (2003). 
\bibitem{singh03} K. Singh, D.M. Tong, K. Basu, J.L. Chen,
and J.F. Du,
Phys. Rev. A   {\bf 67}, 032106 (2003). 
\bibitem{du03} J.F. Du, P. Zou, M. Shi, L.C. Kwek, J.-W. Pan, 
C.H. Oh, A. Ekert, D.K.L. Oi, and M. Ericsson, 
Phys. Rev. Lett. {\bf 91}, 100403 (2003). 
\bibitem{bhandari91} R. Bhandari,
Phys. Lett. A {\bf 157}, 221 (1991);  
Phys. Lett. A {\bf 171}, 262 (1992);
Phys. Lett. A {\bf 171}, 267 (1992);
Phys. Lett. A {\bf 180}, 15 (1993).
\bibitem{bhandari02} R. Bhandari,
Phys. Rev. Lett. {\bf 89}, 268901 (2002);  
J.S. Anandan, E. Sj\"{o}qvist, A.K. Pati,
A. Ekert, M. Ericsson, D.K.L. Oi, and V. Vedral,
Phys. Rev. Lett. {\bf 89}, 268902 (2002).
\bibitem{manini00} N. Manini and F. Pistolesi, 
Phys. Rev. Lett. {\bf 85}, 3067 (2000).
\bibitem{mukunda02} N. Mukunda, Arvind, S. Chaturvedi, and R. Simon,   
Phys. Rev. A {\bf 65}, 012102 (2002). 
\bibitem{hasegawa01} Y. Hasegawa, R. Loidl, M. Baron, G. Badurek, 
and H. Rauch, 
Phys. Rev. Lett. {\bf 87}, 070401 (2001).   
\bibitem{hasegawa02} Y. Hasegawa, R. Loidl, G. Badurek, 
M. Baron, N. Manini, F. Pistolesi, and H. Rauch, 
Phys. Rev. A {\bf 65}, 052111 (2002). 
\bibitem{filipp03} S. Filipp and E. Sj\"{o}qvist, 
Phys. Rev. Lett. {\bf 90}, 050403 (2003). 
\bibitem{sjoqvist03} E. Sj\"{o}qvist and S. Filipp, 
in {\it Foundations of probability and physics - 2},
ed. A. Khrennikov; series ``Math.  Modelling in Physics, 
Engineering and Cognitive Sciences'' (V\"{a}xj\"{o} Univ. Press, 
V\"{a}xj\"{o}, 2003), 545. 
\bibitem{remark1} If the density operator has some degenerate 
eigenvalues, no unique basis is singled out, which makes the 
parallel transport condition ambiguous.  
\bibitem{bures69} D. Bures, 
Trans. Am. Math. Soc. {\bf 135}, 199 (1969); 
A. Uhlmann,
Rep. Math. Phys. {\bf 9}, 273 (1976).
\bibitem{jozsa94} R. Jozsa,  
J. Mod. Opt. {\bf 41}, 2315 (1994). 
\bibitem{wagh95} A.G. Wagh and V.C. Rakhecha, 
Phys. Lett. A {\bf 197}, 107 (1995); 
A.G. Wagh, V.C. Rakhecha, P. Fischer, and A.I. Ioffe, 
Phys. Rev. Lett.{\bf 81}, 1992 (1998).
\bibitem{remark2} Nothing prevents us to consider more complicated 
$F^{(l)}$'s. For example, one may add a function $G^{(l)} (\rho)$ to
$\rho^{p/q}$ provided it converges for any $\rho$ and $G^{(l)} (P) =0$
for any projector $P$. Although considerations of such alternative
definitions may have some mathematical interest, we believe the
present choice is, from a physical point of view, the most natural,
as it turns out to be reducible to the mixed state phase of Ref. 
\cite{sjoqvist00} for $l=1$ and it can be realized experimentally, 
at least for $l\leq 2$.
\bibitem{remark3} Notice that $\rho_{j_k}^{p/q}$ is well-defined 
since $\rho_{j_k} \geq 0$. 
\bibitem{franson89} J.D. Franson,
Phys. Rev. Lett. {\bf 62}, 2205 (1989).
\bibitem{kwiat99} P.G. Kwiat, E. Waks, A.G. White, I. Appelbaum, 
and P.H. Eberhard,
Phys. Rev. A {\bf 60}, R773 (1999); 
A.G. White, D. James, P. Eberhard, and P. Kwiat,
Phys. Rev. Lett. {\bf 83}, 3103 (1999). 
\bibitem{hessmo00} B. Hessmo and E. Sj\"{o}qvist, 
Phys. Rev. A {\bf 62}, 062301 (2000). 
\end{thebibliography}
\end{document}